\documentclass[a4paper]{JHEP3}
\usepackage{epsfig}
\usepackage{amsmath,graphicx}

\def\,{\ifmmode\mskip\thinmuskip\else\leavevmode\thinspace\fi}

\def\ba{\begin{eqnarray}}
\def\ea{\end{eqnarray}}

\def\dd{{\mathrm d}}

\def\fun#1#2{\lower3.6pt\vbox{\baselineskip0pt\lineskip.9pt
  \ialign{$\mathsurround=0pt#1\hfil##\hfil$\crcr#2\crcr\sim\crcr}}}

\def\gsim{\mathrel{\raise.3ex\hbox{$>$\kern-.75em\lower1ex\hbox{$\sim$}}}}
\def\lsim{\mathrel{\raise.3ex\hbox{$<$\kern-.75em\lower1ex\hbox{$\sim$}}}}

\title{Classical Scalar Field Potential in the Standard Model}

\author{Andrej B. Arbuzov,
Lukasz A. Glinka,
Victor N. Pervushin \\
Bogoliubov Laboratory of Theoretical Physics, \\
Joint Institute for Nuclear Research, 141980 Dubna, Russia \\
E-mail: \email{arbuzov@theor.jinr.ru}
}

\abstract{The mechanism of particle mass generation in the Standard Model is discussed.
It is shown that non-zero vacuum expectation value of a scalar field together with the proper
symmetry of the Lagrangian allow a certain class of scalar field potentials providing
generation of gauge boson and fermion masses and keeping the electroweak sector of 
the model unchanged. 
Applying the minimality principle and certain additional conditions
one can reduce the number of free parameters in the model.
Possible phenomenological consequences in the scalar sector for different choices of the 
potential are discussed. }


\preprint{arXiv:08XX.XXXX [hep-ph]}

\begin{document}

\section{Introduction}

The Standard Model (SM) of elementary particles and fundamental interactions
is a very successful physical theory providing theoretical predictions being
in an excellent agreement with practically all present high energy physics experimental
data. Nevertheless, for many reasons we suppose that the SM is not
the ultimate theory of everything, but rather an effective model appropriate for
the given energy range of modern experiments. Moreover, one of the most important
ingredients of the model, the mechanism of electroweak (EW) symmetry breaking and
particle mass generation, has not yet been directly verified.
In the forthcoming experiments, particularly at the Large Hadron Collider (LHC) 
at CERN, we hope to access both the limit of the SM applicability 
and the mechanism of the EW symmetry breaking.

In the SM, the Higgs--Kibble mechanism~\cite{Higgs:1966ev,Kibble:1967sv}
together with the Yukawa interactions of the Higgs scalar with fermions
are responsible for generation of masses for weak bosons and fermions,
respectively. Note that in the SM the mass of the Higgs boson, $m_h$, appear as
a free parameter and it is not {\em generated} contrary to all others.
The direct experimental limit, $m_h>114.4$~GeV at 95\% C.L.,
and indirect upper bounds can be imposed on the mass of the SM Higgs particle
(see {\it e.g.} Refs.~\cite{Yao:2006px,Djouadi:2005gi} and references therein).
But still the nature of the very origin of the Higgs boson are not well
understood and justified contrary to the ones of the gauge bosons, which
are believed to be in a deep relation with the space-time symmetry properties.
Moreover, there are several difficulties in the SM directly related to the
Higgs potential: tachyon behavior of the Higgs field at large energies,
monopole solutions in the classical Higgs potential, non-zero imaginary part
of the effective potential, large Higgs self-coupling leading to non-perturbative
effects and possibly to unitarity violation, and other problems. These difficulties and
especially the naturalness (or the fine tuning) problem motivate us to look for
models beyond SM.
In this context the discussion of possible modifications of the scalar sector
in the SM discussed in this paper might be of interest.

We suggest to consider a generalization of the scalar field potential. 
In fact, to generate the masses of electroweak bosons it is sufficient 
to have a non-zero vacuum expectation value of the scalar field
\ba 
\sqrt{2}\langle 0|\Phi|0 \rangle \equiv \eta =
\left(\sqrt{2}G_{\mathrm{Fermi}}\right)^{-1/2} \approx 246~\mathrm{GeV}
\ea 
together with a certain symmetry condition corresponding to the symmetry in the 
gauge sector of the model. On the other hand the standard Higgs potential with 
a tachyon mass parameter does not follow from any basic principle used 
to construct the SM. In particular the Higgs potential is 
not minimal as will be shown below. Moreover, it uses the correspondence 
to the Ginzburg-Landau superconductivity mechanism, where the appearance 
of such a potential is provided by external conditions, while in the SM 
this correspondence looks artificial since we to try to construct the
model as a fundamental theory resulting from the {\em first principles}.

The paper is organized as follows. In the next Sect. we will consider in detail the
mechanism of mass generation in the $U(1)$ abelian case. The conditions on the
classical potential are discussed in Sect.~\ref{sect_v}.
In Conclusions we discuss the $SU(2)\times U(1)$ case and peculiar 
properties of certain scalar potentials and their possible impact on the
phenomenology.

\section{Mass generation in the abelian case}\label{sect_abel}

Let us start with a model describing interactions of a scalar field $\Phi$ and a vector
abelian gauge field $A$ with the Lagrangian
\begin{eqnarray}\label{lag}
 \mathcal{L} = \partial_{\mu}\Phi^{\dagger}\partial^{\mu}\Phi - V(\Phi)
- \dfrac{1}{4}F_{\mu\nu}F^{\mu\nu}
+ ig\left(\Phi^{\dagger}\partial_{\mu}\Phi-(\partial_{\mu}\Phi^{\dagger})\Phi\right)A^{\mu}
+ g^2\Phi^{\dagger}\Phi A_{\mu}A^{\mu},
\end{eqnarray}
where $g$ is the charge of the scalar field, and $V(\Phi)$ is the classical potential
of the scalar field.
We require the symmetry of the Lagrangian with respect to the transformations
\ba
\Phi \to e^{i\chi}\Phi, \qquad
\Phi^{\dagger} \to e^{-i\chi}\Phi^{\dagger}, \qquad
A_{\mu} \to A_{\mu} + \frac{1}{g}\partial_{\mu}\chi.
\ea
For the potential $V(\Phi)$ the above condition means that it should depend only
on the product $\Phi^{\dagger}\Phi$ which is invariant under these transformations:
\ba\label{phi2}
V(\Phi) = V(\Phi^{\dagger}\Phi).
\ea

The scalar field in the present case possesses two degrees of freedom. We can
use the polar coordinates reflecting the symmetry of the theory 
and cast it in the form
\ba \label{polar}
\Phi(x) = \sigma(x) e^{i\theta(x)}, \qquad
\Phi^{\dagger}(x) = \sigma(x) e^{-i\theta(x)}.
\ea
The standard procedure of the gauge boson mass generation
can be performed in these variables. Substituting (\ref{polar}) in
the Lagrangian~(\ref{lag}) we get
\ba
 \mathcal{L} = \partial_{\mu}\sigma\partial^{\mu}\sigma - V(\Phi)
+ g^2\sigma^2(A_{\mu}+\frac{1}{g}\partial_{\mu}\theta)
(A^{\mu}+\frac{1}{g}\partial^{\mu}\theta)
- \frac{1}{4}F_{\mu\nu}F^{\mu\nu}.
\ea
To get a mass for the gauge boson it is sufficient to suppose that there is
a non-zero vacuum expectation value of the radial degree of freedom of the scalar field:
\ba
\langle0|\sigma |0\rangle \equiv \sigma_0 = \frac{\eta}{\sqrt{2}},
\ea
so that
\ba\label{eta_h}
\sigma(x) = \frac{\eta+h(x)}{\sqrt{2}},
\ea
where $h$ is a usual particle-like excitation which is called the Higgs boson,
$\langle0 |h(x)| 0\rangle = 0$.

Note that the vacuum expectation value of the scalar field can be treated as its
zeroth harmonic corresponding to an average over a large space volume $V_0$:
\ba \label{average}
\frac{1}{V_0}\int_{V_0}\dd^3x\,\sigma(x) = \sigma_0, \qquad
\frac{1}{V_0}\int_{V_0}\dd^3x\, h(x) = 0.
\ea

If we fix the gauge of our vector field as
\ba \label{gauge_fix}
A_\mu(x) \to B_\mu(x) = A_\mu(x) + \frac{1}{g}\partial_{\mu}\theta(x)
\ea 
and apply the separation of the scalar field zeroth harmonic, we get
the Lagrangian in the form
\ba
\mathcal{L} &=& \frac{1}{2}\partial_{\mu}h\partial^{\mu}h 
- V\left(\frac{\eta+h(x)}{\sqrt{2}}\right)
+ \frac{1}{2}g^2\eta^2 B_{\mu}B^{\mu} 
+ g^2\eta h B_{\mu}B^{\mu} + \frac{1}{2}g^2h^2B_{\mu}B^{\mu}
\nonumber \\
&-& \frac{1}{4}\widetilde{F}_{\mu\nu}\widetilde{F}^{\mu\nu}, \qquad \qquad
\widetilde{F}_{\mu\nu}=\partial_\mu B_\nu - \partial_\nu B_\mu.
\ea
So the vector field $B$ acquired a non-zero mass by absorbing the rotational
degree of freedom of the scalar field. Note that this degree of freedom is
massless just because it describes rotation without any reference to the
Goldstone theorem. On the other hand, gauge fixing~(\ref{gauge_fix}) can be
considered as the $U(1)$ symmetry breaking. 
Nevertheless as discussed in Ref.~\cite{Chernodub:2008rz} 
keeping the phenomenology unchanged, the whole 
procedure of mass generation performed in the polar variables 
can be interpreted in terms of supercurrents 
without any explicit gauge fixing.

Generation of the fermion masses can be performed in the usual way
by introducing in the Lagrangian additional terms for free massless
fermions and their interaction with our scalar field. Note that in the
polar variables the Yukawa interaction term can be taken in the $U(1)$ 
symmetric form as
\ba \label{fermion_mass}
g_f|\Phi|\bar{f}f = g_f\sigma\bar{f}f = \frac{g_f\eta}{\sqrt{2}}\bar{f}f
+ \frac{g_f}{\sqrt{2}}h\bar{f}f.
\ea

\section{The scalar field potential}\label{sect_v}

Let us look now at the scalar field potential and define the class of
possible choices of its form.

As concerns the symmetry condition for the potential, in these variables
we explicitly see that it can be a bit extended with respect to the usual one
defined by Eq.~(\ref{phi2}):
\ba\label{phi1}
V(\Phi) = V(|\Phi|) = V(\sigma).
\ea

Cosmological observations show that the Higgs contribution to the Universe
energy density vanishes (or is extremely small). This gives us an additional
condition on the potential:
\ba \label{vv=0}
V(\sigma_0) + V_{\mathrm{eff}}(\sigma_0) = 0, 
\ea
where $V_{\mathrm{eff}}(\sigma_0)$ is the effective Coleman-Weinberg additional
part of the potential coming from loop corrections~\cite{Coleman:1973jx}. 
Note that this condition is a big puzzle for the SM, even so that it can 
be adjusted by {\em tuning} the divergence subtraction procedure in the
$V_{\mathrm{eff}}$ calculations. In what follows we assume that we can choose 
such a classical potential that
\ba \label{v=0}
V(\sigma_0) = 0, \qquad V_{\mathrm{eff}}(\sigma_0) = 0,
\ea
taking into account that a constant shift in the potential corresponds to 
adding a full derivative to the Lagrangian.

One more condition is coming from the stability condition: the point $|\Phi|=\sigma_0$
has to be a minimum of the potential, so that
\ba \label{vpr}
\left. \frac{\dd V(\sigma)}{\dd\sigma}\right|_{\sigma=\sigma_0} = 0, \qquad
\left. \frac{\dd^2 V(\sigma)}{\dd\sigma^2}\right|_{\sigma=\sigma_0} \leq 0.
\ea
This minimum has to be at least a local one. But from the general point
of view it would be much better if that is the unique global minimum
of the potential. 

Let us limit the class of potentials by a polynomials of the 4th order or lower:
\ba\label{v_pol}
V_{\mathrm{pol}}(\sigma) = c_0 + c_1\sigma + c_2\sigma^2 + c_3\sigma^3 + c_4\sigma^4.
\ea  
For the minimality reason we do not consider now so-called quasi-potentials 
and higher order operators. Applying the considered conditions on the coefficients
of the above potential and dropping terms linear in
the Higgs field $h$ because of Eq.~(\ref{average}), 
we get a class of potentials being different in the scalar sector
but leading to the same effect of vector boson and fermion mass generation.

The standard Higgs potential 
\ba
V_{\mathrm{SM}}(\Phi) = \lambda\biggl(\Phi^{\dagger}\Phi-\frac{\eta^2}{2}\biggr)^2 
= \lambda\eta^2h^2 + \lambda\eta h^3 + \frac{\lambda}{4}h^4.
\ea
certainly satisfies the conditions listed above. In this case we have 
one free parameter $\lambda$, and the Higgs
boson mass is defined from the first term on the right hand side, 
$m_h=\eta\sqrt{2\lambda}$. This form of the potential is known to be
the source of many difficulties in the SM. In particular, the presence
of two minima gives rise to an imaginary part in the one-loop effective 
potential leading to instability of quantum states in 
it as discussed in Refs.~\cite{Weinberg:1987vp,Einhorn:2007rv}.

Let us look at a sub-class of the general potential~(\ref{v_pol}) with
a single global minimum and even in powers of $(\sigma-\sigma_0)$:
\ba\label{v_1}
V_{I}(\sigma) = m_h^2(\sigma-\sigma_0)^2 + \lambda(\sigma-\sigma_0)^4 =
\frac{m_h^2}{2}h^2 + \frac{\lambda}{4}h^4,
\ea
where we have two free parameters: the Higgs mass $m_h$ and the
self-coupling $\lambda$. Note that the triple Higgs self-interaction is 
absent in this case. Potential $V_{I}$ has one more free parameter 
with respect to the standard case. But since there is no any relation
between the parameters, one can consider three special cases.

First we can drop the Higgs self-interaction term:
\ba\label{v_m}
V_{II}(\sigma) = m_h^2(\sigma-\sigma_0)^2 = \frac{m_h^2}{2} h^2.
\ea 
In this case the potential is reduced just to a mass term for the
physical Higgs field. The number of free parameters is the same as in the
standard case. Note that Higgs self-interactions will still appear due
to vector boson and fermion loop corrections. 

There is another interesting case, when the mass term is absent in the classical potential:
\ba\label{v_l}
V_{III}(\sigma) = \lambda(\sigma-\sigma_0)^4 =
\frac{\lambda}{4}h^4.
\ea 
This type of potential can naturally appear if we start from a theory
with the conformal symmetry. Breaking of this symmetry than
can be provided by non-zero vacuum expectation values of scalar components
of the theory. 

Moreover, introduction of the mass into the model {\em by hands}
can be avoided just by setting all the free parameters to zero:
\ba\label{v_0}
V_{IV}(\sigma) \equiv 0.
\ea
In the last two cases, the observable Higgs boson mass should be generated by 
a certain additional mechanism and tasking into account radiative corrections.

Certainly the last choice of the Higgs potential is the minimal one. For all
the four choices~(\ref{v_1}---\ref{v_0}) with a single minimum 
we have differences from the standard case only in the Higgs self-interaction
sector of the Lagrangian including the Higgs mass term. Phenomenological 
consequences and theoretical aspects of the different choices have to be studied
and discussed.

\section{Conclusions}\label{sect_concl}

If we take the full electroweak sector of the Standard Model with the
$SU(2)_L\times U(1)_Y$ symmetry, we can generate the masses of
the EW vector bosons and of all fermions in the standard way again
using the polar coordinates (see {\it e.g.} Ref.~\cite{Chernodub:2008rz}).
Again the key conditions are the proper symmetry of the potential and
the non-zero vacuum expectation value.
In the same way as for the abelian case we can generalize the possible
class of the scalar potential to the form~(\ref{v_pol}) and 
consider the possibility to have a concave potential with a single minimum. 

There is a statement~\cite{Luscher:1988gc,Chivukula:1996sn} 
that for the scalar sector of the SM, being a $\Phi^4$ theory,
to remain perturbative at all scales one needs to have the trivial case without
self-interactions of the scalar bosons, {\it i.e.} $\lambda=0$. This possibility
can be accessed for certain choices of the parameters in the generalized potential
as discussed above. Note that in this case the theory would be free at the classical
level from the Higgs self-interactions and contain only the gauge and Yukawa ones.

In this way we suggest to generalize the class of classical potentials of a scalar field, 
which can be used to generate masses of the SM particles. Certainly, different classical
potential lead to different quantum theories. Several problems such as radiative corrections 
to the Higgs boson mass and the Coleman-Weinberg effective potential can be approached for the
suggested potentials by compilation of the existing SM calculations. Moreover,
the choice of the classical potential in a general case should be motivated by 
a certain basic principle of the theory under construction, {\it e.g.} the correspondence
principle or the conformal symmetry. So we have to look for such a motivation.
Discussion of these questions will be presented elsewhere~\cite{inpreparation}.

\acknowledgments
The authors are cordially grateful to B.M.~Barbashov, A.~Borowiec, A.V.~Efremov, 
D.I.~Kazakov, V.B.~Priezzhev,
and D.V.~Shirkov for interest and valuable discussions.
One of us (A.A.) if grateful for the support by the INTAS grant 05-1000008-8328
and by the grant of the RF President (Scientific Schools 3312.2008.2).

\end{document}